# On Uses of Van der Waerden Test: A Graphical Approach

Elsayed A. H. Elamir[1]

Management & Marketing Department, College of Business,
University of Bahrain, P.O. Box 32038,
Kingdom of Bahrain

## Abstract

Although several nonparametric tests are available for testing population identical distributions or equal means in multiple groups problem, the Van der Waerden test has asymptotically the same efficiency as the classical one-way analysis of variance test under normality assumption where it depends on inverse normal score. In this study the Waerden test statistic is used to create a form that assists in deriving adjusted p values chart in terms of Bonferroni and Benjamini-Hochberg methods. The sampling distribution of the proposed form is derived as a gamma distribution that utilized to acquire initial p values. The proposed method will provide well recognition and deep knowledge where the changes occur. Simulation study is executed to study the characteristic of the proposed method in terms of size and power of the test and compared with original Waerden test and one-way analysis of variance test. The results are of great practical values since the proposed method stretches visible shape to decision maker to gives complete picture and deep insight where the differences happen, keep size and power of the test almost similar to original Waerden test and gives a glimpse of various information that completes the pairwise comparisons. To show the benefits of the proposed method, two applications are presented.

**Keywords:** ANOVA, Bonferroni approximation; nonparametric test, nonnormality, test power; type I error.

---

[1]Email: shabib@uob.edu.bh

# 1 Introduction

It has been known that the one-way analysis of variance (ANOVA) is one of the most used tests in experimental science to explore the effect of one or more factors on a given response variable under certain conditions such as normality, homogeneity of variances and data independent; see, Kutner et al. (2004), Romano (2017), Wang, et al. (2017), Yigit and Mendes (2018), Nguyen et al. (2019) and Elamir (2020). These conditions are difficult to satisfy when the actual data are collected and analysed; see, Kutner et al. (2004), Fox and Weisberg (2019) and Wang et al. (2017). Several nonparametric techniques alternatives to parametric version of ANOVA were derived to be used especially with serious violations of ANOVA conditions or in case of availability of ordinal data; see, Gibbons and Chakraborto (2003) and Pereira and Afonso (2019).

One of these alternatives is rank-based inverse normal transform that transforming the distribution of a random variable to looks more normally distributed. The most used rank-based inverse normal transform is expected normal score proposed by Fisher and Yates (1938) that use historical data of sample quantile or fractional rank to approximate the expected normal scores; see, Beasley et al. (2009). The general form for approximated inverse normal score can be written as

$$S_i = \Phi^{-1}\left(\frac{R_i - c}{n - 2c + 1}\right)$$

$R_i$ is the rank of the $i$th case among all data, $\Phi^{-1}$ is the standard normal quantile and $c$ is offset value, For example, $c = 3/8$ is Blom recommended value (Blom, 1958) and $c = 0$ is Waerden recommended value (Waerden, 1952); see Beasly et al (2009).

Although several nonparametric tests are available, the Waerden test has asymptotically the same efficiency as the classical ANOVA test where it depends on normal scores (inverse normal scores). Gibbons and Chakraborti (2003, page 310) mentioned that "This statistic provides the asymptotically optimum test of location when the population of differences is normal". In addition, it protects against violation of normality and homogeneity of variances; see, Hajek (1969) and Dijkstra and Linders (1987). The Waerden test for several $g$ mutually independent samples is a natural extension of Waerden test in case of two-sample problem; see, Dijkstra and Linders (1987) and Mansouri and Chang (1995). The null hypothesis is that

$$H_0: \text{all } g \text{ population distributions are identical}$$

Versus

$$H_1: \text{at least two population distributions are not (identical) same}$$

But when the location model is assumed, the null hypothesis can be written as

$$H_0: \theta_1 = \cdots = \theta_g$$

Versus

$$H_1: \text{At least two are not same}$$

See; Gibbons and Chakraborto (2003).

Recently, Waerden test are used in many applications in different fields. Vilcekova et al. (2017) have used it to investigate dependence of smoke on total volatile organic compounds, particulate matters, temperatures, and humidity in a study about the effect of indoor air quality in houses of Macedonia. Riera et al. (2018) have used biofilm formation to make comparison between quality of substrates (building parts for an artificial reef) to optimize an eco-friendly material to design a new Ars produced by giant 3D printer. The Waerden test used to decide about the differences between the micro-algae communities developed on the different substrates.

In this study, the Waerden test-statistic is used to create an expression that assists in deriving adjusted p values chart using the Bonferroni and Benjamini-Hochberg methods. This chart can be considered as a decision chart. The sampling distribution of the proposed expression is derived as gamma distribution with two parameters, the size of the data and group size that can be used to obtain initial p-values. The proposed method will furnish well understanding and deep insight where the differences in means or distributions occur. Simulation study is executed to study the characteristic of the proposed method in terms of type I error and power of the test and compared with original Waerden test, and one-way analysis of variance test. The results are of important values because the proposed method stretches visible shape to decision maker, profound understanding where the difference occur, keep the level of significance and power of the test like original Waerden test and does not require post-hoc pairwise comparisons. The proposed method is not intended to replace the existence methods but to complete them. Two applications are studied to show the utilities of the proposed method.

Van der Waerden test is reviewed in Section 2. The proposed method and adjusted p values chart are derived in Section 3. The simulation study is carried out in Section 4. Two applications are described in Section 5. Section 6 is devoted for conclusion.

## 2  Van Der Waerden test

Assume that there are $G$ sets each normally distributed with different means $\theta_g$ and common variance $\sigma^2$, $n_g$ the set or group size, and $X_{ig}$, $i = 1,2,\ldots,n_g$, $g = 1,\ldots,G$, the response observations and $n$ the sample size or the total numbers of data in all sets. Let $R_{ig}$ denote to rank of $X_{ig}$ without regard the group, the Waerden test (1952 and 1953) for several samples from $g$ populations can be written as

$$W = \frac{1}{S^2} \sum_{g=1}^{G} n_g \bar{V}_g^2,$$

where

$$S^2 = \frac{1}{n-1} \sum_{g=1}^{G} \sum_{i=1}^{n_g} V_{ig}^2,$$

and

$$V_{ig} = \phi^{-1}\left(\frac{R_{ig}}{n+1}\right), \quad \bar{V}_g = \frac{1}{n_g} \sum_{i=1}^{n_g} V_{ig}$$

$\phi^{-1}(u)$ is the normal quantile (inverse normal score) of $u$ and $\bar{V} = \frac{1}{g}\sum_{g=1}^{G} \bar{V}_g$; See; Dijkstra and Linders (1987) and Mansouri and Chang (1995). Under the null hypothesis of identical distributions or equal locations, $W$ has chi square with $g - 1$ degrees of freedom; see, Gibbons and Chakraborto (2003). In practice, it may reject the null hypothesis if $W > \chi^2(\alpha; df = g - 1)$ and in this case the post-hoc test for multiple pairwise comparisons are given by Conover and Iman (1979) as

$$\left|\frac{V_i}{n_i} - \frac{V_j}{n_j}\right| > t_{(\frac{\alpha}{2}; n-g)} \sqrt{S^2 \frac{n-1-W}{n-k}\left(\frac{1}{n_i} + \frac{1}{n_j}\right)}$$

Where $V_g = \sum_{i=1}^{n_g} V_{ig}$; see, Pohlert (2021).

Since the Waerden test used inverse normal scores, the chi square distribution gives a very good fit for $W$ until in small sample sizes; see, Conover (1999). Lupsen (2017) carried out extended simulation study for two-way layouts to compare several nonparametric tests ("rank transform", "inverse normal transform", "aligned rank transform", "a combination of aligned rank transforms", "inverse normal transform", "Puri and Sen's L statistic", "Van der Waerden" and "Akritas & Brunners ATS") with ANOVA. He used 16 normal and nonnormal distributions with and without equal variances in terms of type I error and power of the test. Although Van der Warden test had shown a little shortage in performance in some situations such as lack of power in the case of multifactorial design, but in general the test had shown the best overall performance.

Figures 1 and 2 show the histograms of $W$ using different settings ($g = 3, 15$ and $n = 30, 300$) and simulated data from normal, $t(df = 2)$ and $\chi^2(df = 1, 2, 4)$ with chi square distribution superimposed. It is clear that the chi square distribution fit the data well for different setting where the heights of the bar follow closely the shape of the line.

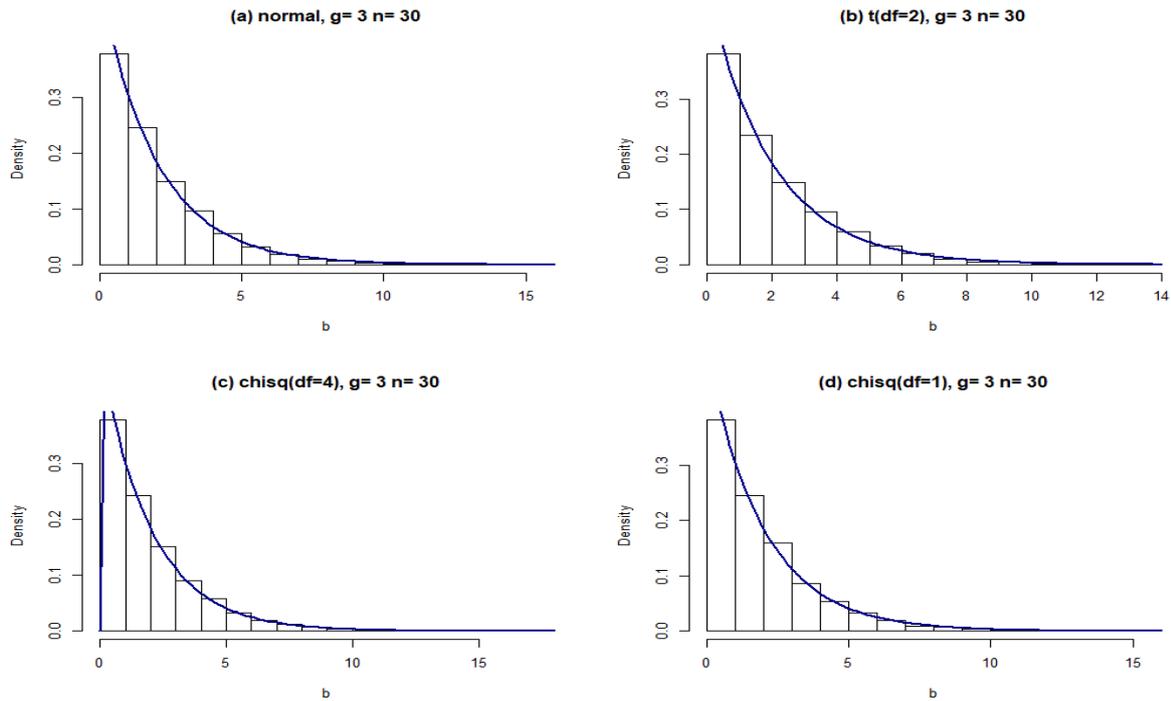

Figure 1 histograms of $W$ using simulated data from different setting with $\chi^2$ distribution superimposed

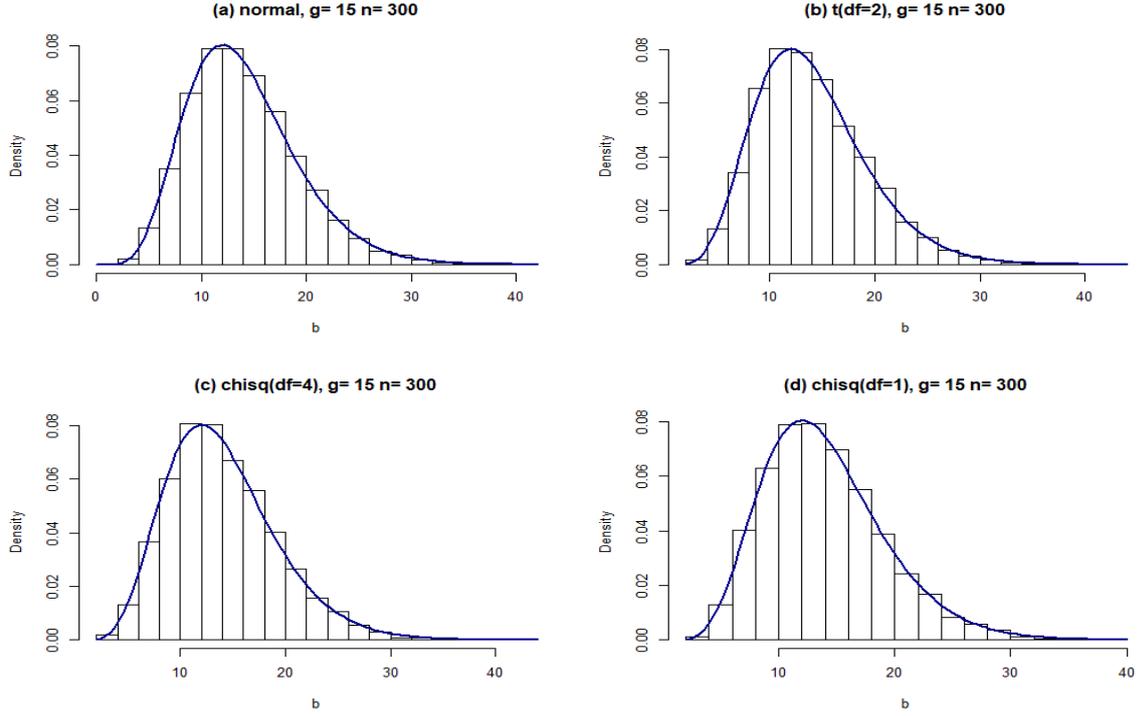

Figure 2 histograms of $W$ using simulated data from different setting awith $\chi^2$ distribution superimposed

## 3 The proposed method and adjusted p values chart

The Van der Waerden test can be written as

$$W = \frac{1}{S^2} \sum_{g=1}^{G} n_g \bar{V}_g^2 = \sum_{g=1}^{G} \frac{n_g \bar{V}_g^2}{S^2} = \sum_{g=1}^{G} E_g$$

The proposed expression to create adjusted p values chart can be written as

$$E_g = \frac{n_g \bar{V}_g^2}{S^2}, \qquad g = 1, \ldots, G$$

This is the ratio of square average of inverse normal scores for each group weighted by group size, $n_g \bar{V}_g^2 = n_g \left[\frac{1}{n_g} \sum_{i=1}^{n_g} V_{ig}\right]^2$, to the variance of inverse normal scores for all treatments (combined samples) $S^2 = \frac{1}{n-1} \sum_{g=1}^{G} \sum_{i=1}^{n_g} (V_{ig})^2$. The sampling distribution of $E_g$ is needed to obtain the initial p values and this can be derived as follows.

### 3.1 The sampling distribution of $E_g$

Since the groups are assumed independent and identically distributed as normal, the sampling distribution of $E_g$ can be derived as

$$E_g = \frac{n_g \bar{V}_g^2}{S^2} = \frac{n_g(\bar{V}_g - \bar{V})^2}{S^2} - \frac{n_g n \bar{V}^2}{nS^2} \sim \left(\frac{n-n_g}{n}\right)\chi^2(1) \sim \Gamma\left(\text{shape} = 0.5, \text{scale} = \frac{2(n-n_g)}{n}\right)$$

By using the density function of gamma distribution,

$$f(x) = \frac{\beta^\alpha}{\Gamma(\alpha)} x^{\alpha-1} e^{-\beta x}$$

The sampling distribution of $E_g$ is derived as

$$f_{E_i}(u) = \frac{(n/2(n-n_g))^{1/2}}{\Gamma(1/2)} u^{-1/2} e^{-\frac{n}{2(n-n_g)}u}, \quad u > 0$$

This is defined in terms of $n$ and $n_g$ with moments

$$E(E_g) = \frac{n - n_g}{n}, \quad Var(E_g) = \frac{2(n-n_g)^2}{n^2}, \quad Sk = 2\sqrt{2}, \text{ and } Ku = 15$$

In case of all the groups have equal size $n_g$, the density function can be re-written as

$$f_{E_g}(u) = \frac{(g/2(g-1))^{1/2}}{\Gamma(1/2)} u^{-1/2} e^{-\frac{g}{2(g-1)}u}, \quad u > 0$$

That is defined in terms of the number of groups $g$.

To see how the gamma distribution is a good approximation to $E_g$ specially for nonnormal distributions, a simulation study is conducted to compute the first four empirical moments of $E_g$ at $g = 3$ and 8, $n_g = 10$ and 25 from normal, Laplace and chi square distributions. The steps can be summarized as

1. Choose the desired design such as $g = 5$, $n_g = 25$, chi square (df=2) distribution,
2. Simulate values from the chosen distribution with equal means,
3. Calculate $E_g, g = 1, \ldots, G$ for each set,
4. Compute the first four moments for each $E_g, g = 1, \ldots, G$,
5. Repeat this many times as required and calculate the average for each setting.

The first four empirical moments for $E_g$ are given in Table 1 at $g = 3$ and 5 and $n_g = 10$ and 25 using normal, Laplace and chi square distributions. Table 1 illustrates

1. the empirical mean and variance are very close to theoretical values for all setting,
2. As the total number of observations, $n$, become large, the empirical skewness and kurtosis are getting closer to theoretical values,
3. In general, since the estimation of kurtosis is unstable, the gamma distribution gives a very good fit for $E_g, g = 1, \ldots, G$.

Table 1 the first four empirical and theoretical moments of $E_g$ using different setting with equal means and number of replications 10000.

| $g$ | $n_i$ | | normal | | | | Laplace | | | | $\chi^2(2)$ | | | |
|---|---|---|---|---|---|---|---|---|---|---|---|---|---|---|
| | | | Mean | Var. | Sk. | Ku. | mean | Var. | Sk. | Ku. | mean | Var. | Sk. | Ku. |
| 3 | 10 | $E_1$ | 0.663 | 0.806 | 2.42 | 10.46 | 0.665 | 0.809 | 2.51 | 11.77 | 0.672 | 0.807 | 2.45 | 11.32 |
| | | $E_2$ | 0.675 | 0.837 | 2.39 | 10.62 | 0.677 | 0.863 | 2.50 | 11.45 | 0.652 | 0.781 | 2.36 | 10.21 |
| | | $E_3$ | 0.671 | 0.817 | 2.42 | 11.57 | 0.676 | 0.839 | 2.48 | 11.78 | 0.657 | 0.815 | 2.63 | 12.93 |
| | | **Theo.** | **0.667** | **0.889** | **2.82** | **15.00** | **0.667** | **0.889** | **2.82** | **15.00** | **0.667** | **0.889** | **2.82** | **15.00** |
| | 25 | $E_1$ | 0.671 | 0.842 | 2.51 | 11.72 | 0.674 | 0.849 | 2.51 | 11.48 | 0.662 | 0.866 | 2.87 | 15.74 |
| | | $E_2$ | 0.665 | 0.844 | 2.69 | 14.12 | 0.676 | 0.895 | 2.74 | 14.10 | 0.659 | 0.827 | 2.59 | 12.24 |
| | | $E_3$ | 0.667 | 0.842 | 2.63 | 12.51 | 0.673 | 0.868 | 2.58 | 11.91 | 0.672 | 0.865 | 2.65 | 13.33 |
| | | **Theo.** | **0.667** | **0.889** | **2.82** | **15.00** | **0.667** | **0.889** | **2.82** | **15.00** | **0.667** | **0.889** | **2.82** | **15.00** |
| 5 | 10 | $E_1$ | 0.802 | 1.239 | 2.60 | 12.46 | 0.808 | 1.22 | 2.59 | 12.01 | 0.785 | 1.17 | 2.64 | 13.11 |
| | | $E_2$ | 0.809 | 1.252 | 2.58 | 11.94 | 0.809 | 1.26 | 2.70 | 13.35 | 0.788 | 1.16 | 2.48 | 11.54 |
| | | $E_3$ | 0.820 | 1.278 | 2.70 | 13.88 | 0.820 | 1.27 | 2.75 | 13.88 | 0.795 | 1.15 | 2.43 | 10.86 |
| | | $E_4$ | 0.804 | 1.237 | 2.62 | 12.42 | 0.804 | 1.16 | 2.62 | 12.42 | 0.804 | 1.16 | 2.62 | 12.42 |
| | | $E_5$ | 0.799 | 1.211 | 2.68 | 11.89 | 0.799 | 1.17 | 2.56 | 12.09 | 0.799 | 1.16 | 2.48 | 11.32 |
| | | **Theo.** | **0.80** | **1.28** | **2.82** | **15** | **0.80** | **1.28** | **2.82** | **15** | **0.80** | **1.28** | **2.82** | **15** |
| | 25 | $E_1$ | 0.79 | 1.19 | 2.84 | 15.22 | 0.80 | 1.23 | 2.61 | 12.28 | 0.79 | 1.23 | 2.77 | 14.75 |
| | | $E_2$ | 0.78 | 1.21 | 2.85 | 15.06 | 0.80 | 1.25 | 2.69 | 13.18 | 0.79 | 1.24 | 2.81 | 14.55 |
| | | $E_3$ | 0.80 | 1.24 | 2.73 | 14.03 | 0.80 | 1.24 | 2.75 | 14.37 | 0.79 | 1.24 | 2.82 | 14.96 |
| | | $E_4$ | 0.78 | 1.20 | 2.85 | 15.30 | 0.79 | 1.22 | 2.70 | 13.18 | 0.78 | 1.16 | 2.65 | 12.74 |
| | | $E_5$ | 0.81 | 1.26 | 2.74 | 14.44 | 0.79 | 1.22 | 2.65 | 12.84 | 0.81 | 1.29 | 2.70 | 13.18 |
| | | **Theo.** | **0.80** | **1.28** | **2.82** | **15** | **0.80** | **1.28** | **2.82** | **15** | **0.80** | **1.28** | **2.82** | **15** |

\* var: variance, sk: skewness, ku: kurtosis, norm: normal, $t$: t-distribution and $\chi^2$: chi square distribution, Theo: theoretical

### 3.2 Adjusted p value chart

To create adjusted p values chart, it has to consider $g$ tests that required making difference between two types of $\alpha$ errors; see, Abdi (2007) and Bretz (2011), as follows.
1. The first one called "experiment alpha or alpha per family" that denoted by $\alpha[PF]$. This is the whole experiment level of significance,
2. The second one called "test-wise or alpha per test" that denoted by $\alpha[PT]$. This is a specific test level of significance.

Finding the first type of error ($\alpha[PF]$) for $G$ tests can be obtained from Abdi (2007) as

$$\alpha(PF) = 1 - (1 - \alpha(PT))^G$$

Then,

$$\alpha(PT) = 1 - (1 - \alpha(PF))^{1/G}$$

The Bonferroni approximation gives a simpler form as

$$\alpha(PT) \approx \frac{\alpha(PF)}{k}$$

In addition to Bonferroni approximation, there are different methods that adjust p values in cases of $g$ tests, such as Holm's method (1979), Simes-Hochberg method (Simes 1986, Hochberg 1988), Hommel's method (1988) and Benjamini-Hochberg method (1995).

Recently, Benjamini-Hochberg method (HB method) gains popular using in adjusting p values. This method control what is called the false discovery rate that is defined as the probability of wrong rejection of a hypothesis happens. HB method is used a sequential way for adjusted Bonferroni correction for many tests instead of the family wise rate; for details about this method; see, Benjamini and Hochberg (1995). Fortunately, HB method and other methods are available in R-software under the name "p.adjust(p; method = (""); n = length(p))" ; see, R Core Team (2021). Therefore, the adjusted p values chart can be proposed as follows.
1. Compute $E_g, g = 1, ..., G$ from data,
2. Calculate initial p values using sampling distribution, i.e, pgamma$(E_g, 0.5, \text{rate} = n/(2(n - n_g)))$,
3. Find adjusted p values using Bonferroni and Benjamini-Hochberg methods or any other method,
4. Plot $g$ against adjusted p values and take decision based on nominal $\alpha$. Hence, the adjusted p values chart is
$$x = 1:g \quad \text{versus} \quad y = \text{adjusted } p \text{ values}$$
5. Take decision about hypothesis as
$$\text{if any adjusted } p \text{ value} < \text{nominal } \alpha, \text{ for } g = 1, ..., G$$
$H_0$ is rejected and this will characterize where the shifts occur.

## 4 Simulation study

Adjusted p values chart using Bonferroni approximation and BH method is compared with Waerden and ANOVA tests in terms of Type I error $p(\text{reject } H_0 | H_0 \text{ is true})$ and power of the test $p(\text{reject } H_0 | H_0 \text{ is false})$ =1- $p(\text{accept } H_0 | H_0 \text{ is false})$=1-type II error.

With respect to Type I error, the following steps are used in simulation:
1. Define the desired setting of the study. In this study $g = 3, 6$, $n_g = 10, 20, 50$ and nominal $\alpha = 0.01, 0.05$,
2. Simulate data from a desired distribution with equal means. The normal distribution as original distribution, $t(df = 1)$ as symmetric heavy-tailed distribution, $\chi^2(df = 2)$ and lognormal distributions as non-symmetric heavy-tailed distributions are used,
3. Obtain $E_g$ and initial p values using gamma distribution,
4. Compute adjusted p values using Bonferroni and HB methods for each setting,
5. Define a dummy variable by giving 1 for reject and 0 else,
6. Repeat the number of required times (R=10000) and compute the average for each setting.

The results for these procedures are given in Table 2. It can conclude that
1. ANOVA gives slightly better results than Waerden test and $E_g$-BH in case of normal distribution and small sample size, while the results are similar with increasing $n$,
2. Waerden and $E_g$-BH tests give better results than ANOVA in case of $t$(df=1) and lognormal distributions (heavy tailed distributions),
3. $E_g$-BH gives slightly better results than $E_g$-Bonf in case of normal distribution while it makes a good improvement in empirical type one error in other distributions,
4. $E_g$-BH and W give almost similar results across all used distributions,

In general, W and $E_g$ BH tend to have adequate type I error control across all used distribution shapes and better results than ANOVA in heavy tailed distributions.

Table 2 empirical type I error based on $E_g$ Bonferroni (Bonf.), $E_g$ BH, Van der Waerden (W), ANOVA, nominal $\alpha = 0.05, 0.01$ using normal (N), chis quare, $t$ and lognormal (LN) distributions and the number of replications is 10000

| | | | N(0; 1) | | | | chisq(1) | | | |
|---|---|---|---|---|---|---|---|---|---|---|
| k | $n_i$ | $\alpha$ | $E_g$ Bonf. | $E_g$ BH | W | ANOVA | $E_g$ Bonf. | $E_g$ BH | W | ANOVA |
| 3 | 10 | 0.05 | 0.041 | 0.043 | 0.045 | 0.049 | 0.045 | 0.048 | 0.052 | 0.041 |
| | | 0.01 | 0.0065 | 0.0067 | 0.0078 | 0.0120 | 0.005 | 0.006 | 0.005 | 0.005 |
| | 20 | 0.05 | 0.045 | 0.047 | 0.051 | 0.054 | 0.042 | 0.045 | 0.048 | 0.043 |
| | | 0.01 | 0.0091 | 0.0093 | 0.0096 | 0.0121 | 0.008 | 0.008 | 0.008 | 0.007 |
| | 50 | 0.05 | 0.043 | 0.046 | 0.049 | 0.051 | 0.043 | 0.046 | 0.049 | 0.046 |
| | | 0.01 | 0.0082 | 0.0085 | 0.0083 | 0.0093 | 0.008 | 0.008 | 0.009 | 0.008 |
| 6 | 10 | 0.05 | 0.045 | 0.047 | 0.043 | 0.051 | 0.042 | 0.043 | 0.042 | 0.042 |
| | | 0.01 | 0.0061 | 0.0061 | 0.0050 | 0.0087 | 0.008 | 0.008 | 0.007 | 0.011 |
| | 20 | 0.05 | 0.046 | 0.048 | 0.048 | 0.053 | 0.045 | 0.046 | 0.045 | 0.042 |
| | | 0.01 | 0.0074 | 0.0080 | 0.0074 | 0.0098 | 0.009 | 0.009 | 0.008 | 0.009 |
| | 50 | 0.05 | 0.049 | 0.050 | 0.048 | 0.050 | 0.045 | 0.047 | 0.050 | 0.049 |
| | | 0.01 | 0.0084 | 0.0084 | 0.0072 | 0.0087 | 0.009 | 0.009 | 0.009 | 0.008 |
| | | | t (df=1) | | | | LN(0, 1) | | | |
| 3 | 10 | 0.05 | 0.037 | 0.040 | 0.041 | 0.014 | 0.039 | 0.042 | 0.045 | 0.033 |
| | | 0.01 | 0.006 | 0.007 | 0.007 | 0.001 | 0.005 | 0.005 | 0.005 | 0.005 |
| | 20 | 0.05 | 0.041 | 0.044 | 0.047 | 0.016 | 0.042 | 0.045 | 0.047 | 0.037 |
| | | 0.01 | 0.007 | 0.007 | 0.007 | 0.002 | 0.007 | 0.008 | 0.008 | 0.006 |
| | 50 | 0.05 | 0.040 | 0.043 | 0.047 | 0.016 | 0.044 | 0.047 | 0.049 | 0.044 |
| | | 0.01 | 0.008 | 0.008 | 0.009 | 0.0013 | 0.009 | 0.010 | 0.009 | 0.006 |
| 6 | 10 | 0.05 | 0.044 | 0.045 | 0.044 | 0.018 | 0.042 | 0.043 | 0.042 | 0.038 |
| | | 0.01 | 0.006 | 0.007 | 0.006 | 0.002 | 0.006 | 0.006 | 0.006 | 0.007 |
| | 20 | 0.05 | 0.043 | 0.045 | 0.048 | 0.017 | 0.043 | 0.045 | 0.043 | 0.038 |
| | | 0.01 | 0.009 | 0.009 | 0.008 | 0.0013 | 0.008 | 0.008 | 0.009 | 0.007 |
| | 50 | 0.05 | 0.047 | 0.048 | 0.052 | 0.017 | 0.045 | 0.047 | 0.047 | 0.041 |
| | | 0.01 | 0.01 | 0.01 | 0.009 | 0.002 | 0.009 | 0.009 | 0.009 | 0.009 |

For the power of the test, the following steps are used in simulation:
1. define the desired setting $g = 3, 6$, $n_g = 10, 20, 50$ and nominal $\alpha = 0.01, 0.05$,
2. Simulate data from a desired distribution with unequal means. The used distributions are the normal distribution with (mean=(1,2,3), sd=2) and (mean=(0,1,2,3,4,5), sd=3), Laplace distribution with (mean=(0,1,2), scale=1) and (mean=(0,1,2,3,4,5), scale=3), chi square distribution with (df=(1,2,4)) and (df=(1,2,3,4,5,6)) and lognormal distribution with ((0,1,2),2) and ((0,1,2,3,4,5), 3),
3. Obtain $E_g$ and compute adjusted p values using Bonferroni and HB methods for each setting,
4. Create a dummy variable by giving 1 for reject and 0 else,
5. Repeat R times and compute the average for each setting.

The results of the test powers are given in Table 3. It can conclude that
1. ANOVA has high power in normal data and less power in lognormal data,
2. The difference in power between $E_g$-BH and W is marginally small in all settings except for Laplace distribution ($g = 6$ and $n_g = 10$),
3. For normal data, the difference in powers among all methods disappear as the sample size increase,
4. $E_g$-BH has slightly better power than $E_g$-Bonf.,
5. The power of $E_g$-BH, $E_g$-Bonf and W is much better than ANOVA using lognormal distribution.

Table 3 empirical power of the test for $E_g$ Bonferroni (Bonf.), $E_g$ BH, Van der Waerden (W), ANOVA, nominal $\alpha = 0.01, 0.05$ using normal (N), Laplace, chi square and lognormal distributions and the number of replications is 10000

| g | $n_i$ | $\alpha$ | $E_g$ Bonf. | $E_g$ BH | W | ANOVA | $E_g$ Bonf. | $E_g$ BH | W | ANOVA |
|---|---|---|---|---|---|---|---|---|---|---|
| | | | \multicolumn{4}{c}{N ((1,2,3), 2)} | \multicolumn{4}{c}{$\chi^2$(df=1,2,3)} |
| 3 | 10 | 0.05 | 0.38 | 0.40 | 0.42 | 0.45 | 0.61 | 0.63 | 0.65 | 0.49 |
| | | 0.01 | 0.15 | 0.15 | 0.16 | 0.22 | 0.32 | 0.33 | 0.35 | 0.23 |
| | 20 | 0.05 | 0.75 | 0.76 | 0.77 | 0.79 | 0.94 | 0.95 | 0.95 | 0.83 |
| | | 0.01 | 0.50 | 0.51 | 0.53 | 0.57 | 0.80 | 0.81 | 0.82 | 0.60 |
| | 50 | 0.05 | 0.99 | 0.99 | 0.99 | 0.99 | 1 | 1 | 1 | 1 |
| | | 0.01 | 0.97 | 0.97 | 0.98 | 0.98 | 0.99 | 0.99 | 1 | 0.98 |
| | | | \multicolumn{4}{c}{N ((0,1,2,3,4,5), 3)} | \multicolumn{4}{c}{$\chi^2$(df=1,2,3,4,5,6)} |
| 6 | 10 | 0.05 | 0.75 | 0.78 | 0.90 | 0.92 | 0.98 | 0.99 | 0.99 | 0.98 |
| | | 0.01 | 0.42 | 0.43 | 0.69 | 0.77 | 0.87 | 0.88 | 0.98 | 0.91 |
| | 20 | 0.05 | 0.99 | 0.99 | 0.99 | 0.99 | 0.99 | 0.99 | 0.99 | 0.99 |
| | | 0.01 | 0.93 | 0.94 | 0.99 | 0.99 | 0.99 | 0.99 | 0.99 | 0.99 |
| | 50 | 0.05 | 1 | 1 | 1 | 1 | 1 | 1 | 1 | 1 |
| | | 0.01 | 1 | 1 | 1 | 1 | 1 | 1 | 1 | 1 |
| | | | \multicolumn{4}{c}{Laplace ((0,1,2), 1)} | \multicolumn{4}{c}{Lognormal ((0,1,2), 2)} |
| 3 | 10 | 0.05 | 0.77 | 0.78 | 0.80 | 0.78 | 0.39 | 0.40 | 0.42 | 0.12 |
| | | 0.01 | 0.48 | 0.50 | 0.52 | 0.55 | 0.15 | 0.15 | 0.16 | 0.03 |
| | 20 | 0.05 | 0.98 | 0.98 | 0.99 | 0.97 | 0.75 | 0.76 | 0.77 | 0.23 |
| | | 0.01 | 0.91 | 0.91 | 0.92 | 0.90 | 0.50 | 0.50 | 0.52 | 0.07 |
| | 50 | 0.05 | 1 | 1 | 1 | 1 | 0.99 | 0.99 | 0.99 | 0.50 |
| | | 0.01 | 1 | 1 | 1 | 1 | 0.97 | 0.97 | 0.97 | 0.25 |
| | | | \multicolumn{4}{c}{Laplace ((0,1,2,3,4,5),3)} | \multicolumn{4}{c}{Lognormal ((0,1,2,3,4,5), 3)} |
| 6 | 10 | 0.05 | 0.51 | 0.54 | 0.65 | 0.62 | 0.74 | 0.77 | 0.89 | 0.11 |
| | | 0.01 | 0.22 | 0.23 | 0.37 | 0.36 | 0.40 | 0.41 | 0.69 | 0.03 |
| | 20 | 0.05 | 0.90 | 0.91 | 0.96 | 0.92 | 0.99 | 0.99 | 0.99 | 0.20 |
| | | 0.01 | 0.67 | 0.67 | 0.86 | 0.80 | 0.92 | 0.93 | 0.99 | 0.07 |
| | 50 | 0.05 | 1 | 1 | 1 | 0.99 | 1 | 1 | 1 | 0.39 |
| | | 0.01 | 0.99 | 0.99 | 0.99 | 0.97 | 1 | 1 | 1 | 0.19 |

# 5 Applications

## 5.1 Application 1

The data for the first application is shown in Table 4 where these data are simulated from five normal distributions with means (100,110,115,100,100) and standard deviation 10. The data consists of 5 groups and the number of observations in each group are 15, 11, 12, 14, 14.

Figure 3 shows the adjusted p values chart for simulated data given in Table 4. This chart is obtained as following

1. The values of $E_g = \frac{n_g \overline{V}_g^2}{S^2}, g = 1, \ldots, 5$ are (1.12, 1.91, 7.18, 0.46 and 3.72),

2. The initial p values are obtained from $\text{pgamma}\left(E_g, \text{shape} = 0.5, \text{rate} = \frac{n}{2(n-n_g)}\right)$ as (0.229, 0.130, 0.003, 0.443, 0.030),

3. The adjusted p values using BH method are obtained using adjusted p function in R-software as "p.adjust(p; method = ("BH"); n = length(p))" as (0.286, 0.217, 0.015, 0.443, 0.074).

Table 4 simulated data from five normal distribution with (means=(100, 110, 115, 100, 100), sd=10)

| A | B | C | D | E |
|---|---|---|---|---|
| 116.23 | 118.98 | 107.76 | 109.32 | 106.19 |
| 115.00 | 116.60 | 103.56 | 82.00 | 87.85 |
| 92.50 | 103.48 | 122.40 | 102.65 | 98.11 |
| 103.77 | 115.51 | 104.06 | 100.01 | 96.91 |
| 98.40 | 94.23 | 120.97 | 116.91 | 106.21 |
| 113.21 | 132.91 | 100.96 | 105.74 | 100.72 |
| 103.35 | 84.68 | 107.59 | 101.32 | 92.65 |
| 98.41 | 106.33 | 113.46 | 99.63 | 96.23 |
| 106.14 | 103.84 | 122.04 | 85.89 | 114.46 |
| 104.45 | 118.25 | 110.30 | 107.19 | 113.54 |
| 81.44 | 110.28 | 126.45 | 88.34 | 88.74 |
| 104.63 | | 123.04 | 114.56 | 88.87 |
| 113.95 | | | 115.18 | 98.78 |
| 74.16 | | | 109.11 | 102.54 |
| 104.23 | | | | |

The results adjusted p values chart using BH and Bonferroni methods are displayed in Figure 3. In both charts, since the adjusted p value for third group C is 0.015 is less than 0.05, the hypothesis of equal means is rejected. Figure 3 is not pairwise comparisons, but it gives complete picture and deep insight where the differences are happening. For example, it can conclude that the group C has the most contribution (7.18/14.38=50%) for causing difference, followed by group E (3.72/14.38=0.26%), group B (1.91/14.38=13%), group A (1.12/14.38=8%), then group D (0.46/14.38=3%) where $E_g = (1.12, 1.91, 7.18, 0.46, 3.72)$ and $\sum_{g=1}^{5} E_g = 14.38$.

On the other hand, the results of original Waerden test are given in Table 5. Since the value of Waerden statistics is 14.388 with p-values 0.0061, the null hypothesis of identical distributions or equal means are rejected without any information about where the differences occur. Also, the table shows the results of post hoc analysis using R package PMCMRplus; see, Pohlert (2021). Table 5 shows the difference occurred mostly between group C-D and group C-E. This may give indication that the group C causes the difference.

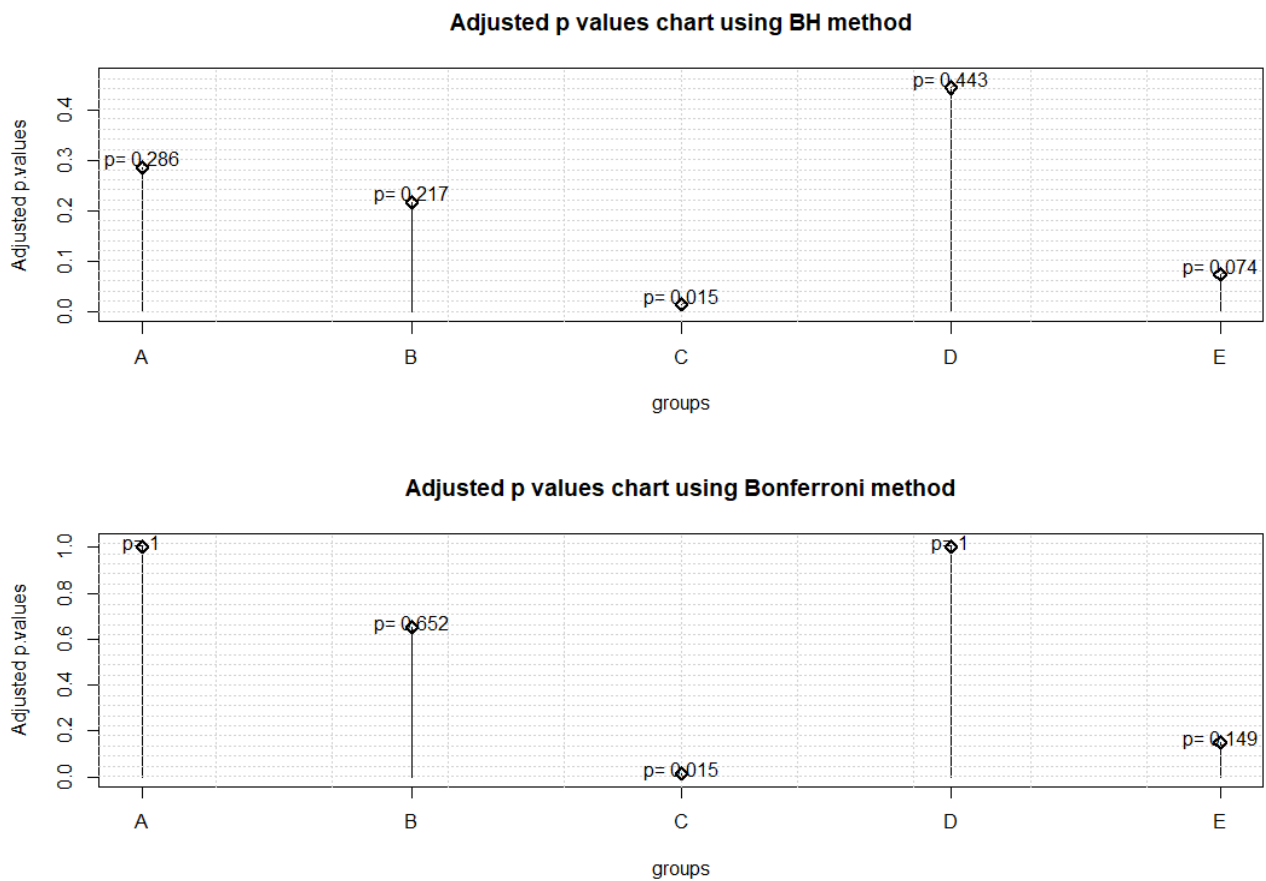

Figure 3 adjusted p values chart using BH and Bonferroni methods

Table5 Waerden test and adjusted p values (BH method) using post hoc analysis for simulated data

|  | Waerden |  | Post hoc |  |  |  |
|---|---|---|---|---|---|---|
| Statistic | 14.388 |  | A | B | C | D |
| P value | 0.0061 | B | 0.122 | - | - | - |
|  |  | C | 0.021 | 0.439 | - | - |
|  |  | D | 0.788 | 0.180 | 0.033 | - |
|  |  | E | 0.529 | 0.034 | 0.0065 | 0.439 |

## 5.2 Application 2

In agricultural experimental sections, the numbers of insects treated with 6 types of different insecticides spray are given in Table 6; see, Beall (1942) and McNeil (1977). The data consists of one factor with 6 different types of spray and 12 counts of insects in each type as response variable. The objective is to test for differences in means or identical distributions.

Table 6 Counts of insects using 6 types of insecticides spray

| Type of spray | | | | | |
|---|---|---|---|---|---|
| A | B | C | D | E | F |
| 10 | 11 | 0 | 3 | 3 | 11 |
| 7 | 17 | 1 | 5 | 5 | 9 |
| 20 | 21 | 7 | 12 | 3 | 15 |
| 14 | 11 | 2 | 6 | 5 | 22 |
| 14 | 16 | 3 | 4 | 3 | 15 |
| 12 | 14 | 1 | 3 | 6 | 16 |
| 10 | 17 | 2 | 5 | 1 | 13 |
| 23 | 17 | 1 | 5 | 1 | 10 |
| 17 | 19 | 3 | 5 | 3 | 26 |
| 20 | 21 | 0 | 5 | 2 | 26 |
| 14 | 7 | 1 | 2 | 6 | 24 |
| 13 | 13 | 4 | 4 | 4 | 13 |

Source: Pohlert, (2021)

Figure 4 shows the boxplot for the data. The distributions of the data are clearly different in location where the median for groups (C, D, and E) is considerably less than the other groups. By noting the median, the distributions of the data are skewed.

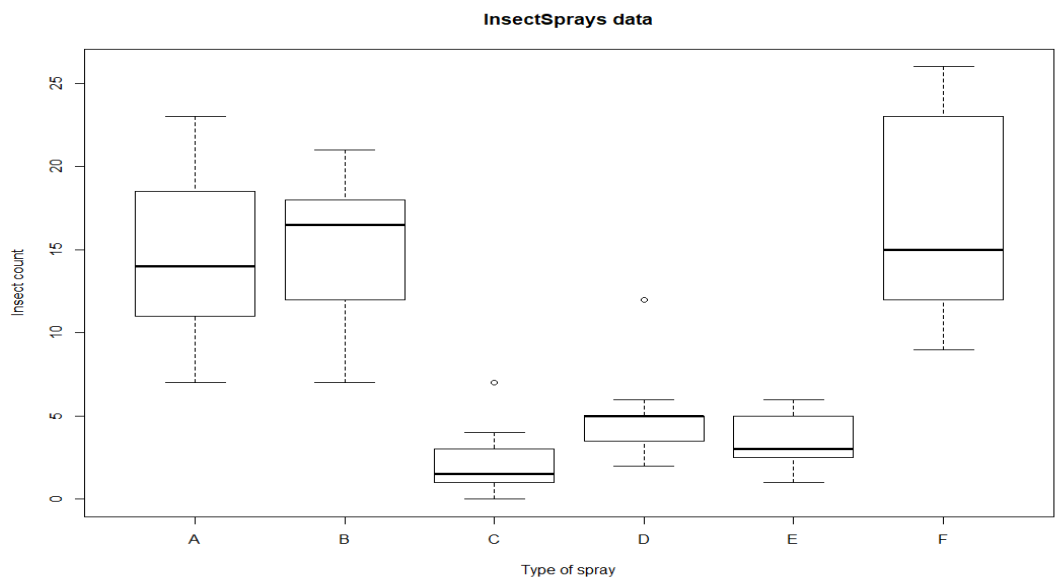

Figure 4 boxplot for number of insects based on 6 types of insect spray

Table 7 shows the results of Waerden test and the post hoc pairwise comparisons for insect spray data. The test is significance and most of pairwise comparisons are significant at levels 0.01 and 0.05 except (A-B, A-F, B-F and D-E)

Table 7 Waerden test and p adjusted (BH method) for post hoc pairwise comparisons for insect spray data

| Waerden | | | Post hoc | | | | |
|---|---|---|---|---|---|---|---|
| statistic | 50.302 | | A | B | C | D | E |
| P value | 0.000000012 | B | 0.636 | - | - | - | - |
| | | C | 0.000 | 0.000 | - | - | - |
| | | D | 0.000015 | 0.000 | 0.0012 | - | - |
| | | E | 0.000 | 0.000 | 0.0431 | 0.2399 | - |
| | | F | 0.268 | 0.501 | 0.000 | 0.000 | 0.000 |

Figure 5 displays the results of adjusted values chart using BH and Bonferroni methods. In both charts, since the adjusted p value for groups B, C, E and F is less than 0.01 and 0.05, this gives strong indication for not identical distributions. Moreover, it can conclude that the group C has the most contribution (17.6/50.3=35%) for causing difference, followed by group F (10.5/50.3=0.21%), group B (7.3/50.3=14.5%), group E (6.9/50.3=13.8%), group A (5.4/50.3=10.7%) and D (2.04/50.3=4%) where $E_g = (5.4, 7.3, 17.6, 2.04, 6.96, 10.47)$ and $\sum_{g=1}^{6} E_g = 50.3$.

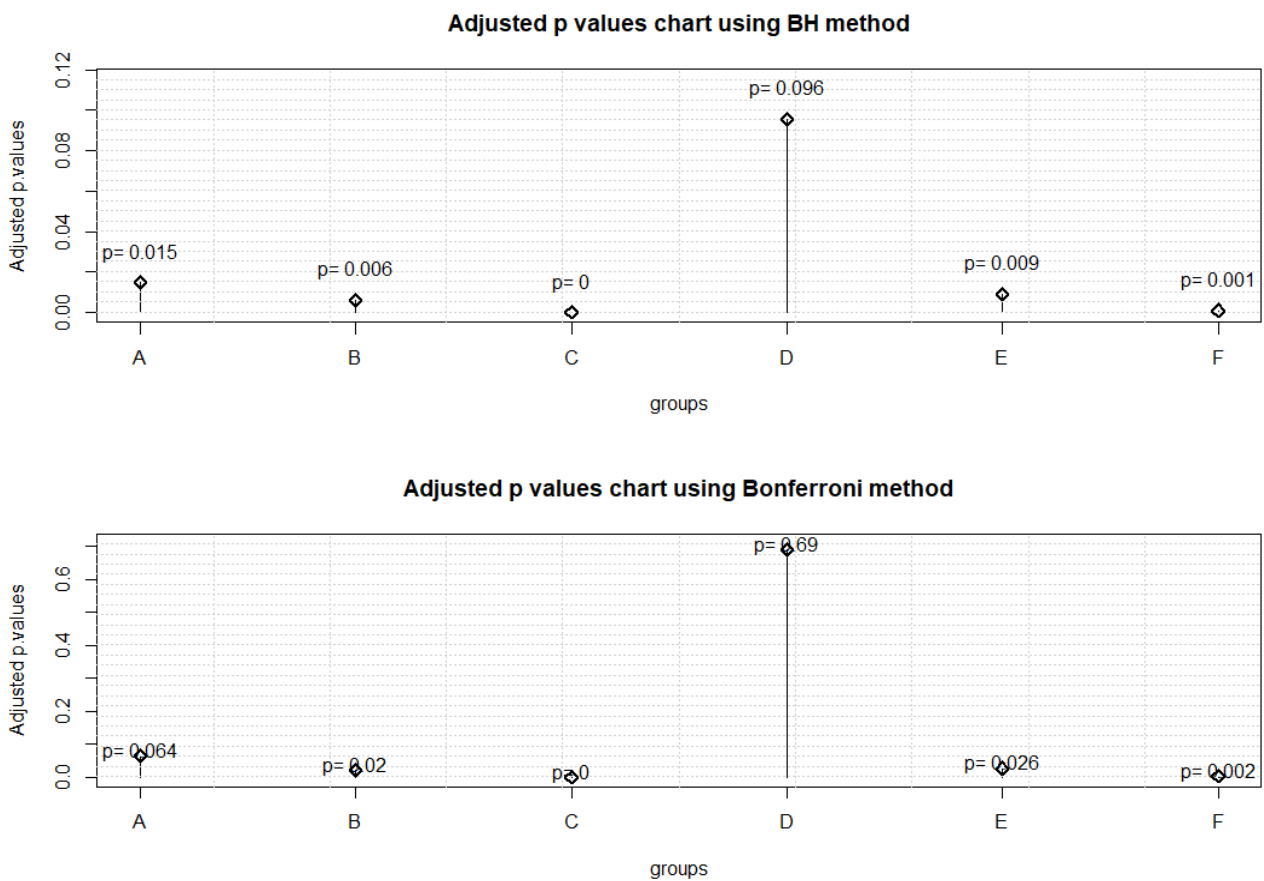

Figure 5 adjusted p values chart for insect spray data (count of insects)

# 6 Conclusion

When the parent distribution of the data is normal, the Waerden test provides asymptotically optimum test of location and had been used in many areas for assessing the hypothesis of identical distributions or equal means. Adjusted p values chart for Waerden test is proposed as the ratio of average square of inverse normal scores for each group weighted by group size to the variance of inverse normal scores for all treatments where the total of $E_g$ is the original Waerden test. The sampling distribution of $E_g$ is derived as a gamma distribution in terms of total number of data and group size. Bonferroni approximation and Benjamini-Hochberg methods were used to obtain adjusted p values in order to decide about identical distributions or equal means when all adjusted p values more than nominal value or not identical distributions when any value of adjusted p values is less than nominal value.

The simulation results illustrated that the performance of adjusted p values chart is similar to Waerden test in terms of type I error and power of the test using different setting. For type one error, ANOVA and Waerden test had given similar empirical results in case of normal distribution and Waerden and $E_g$-BH tests had shown better results than ANOVA in case of heavy tailed distributions ($t$(df=1) and lognormal distributions). Also, $E_g$-BH clarified slightly better results than $E_g$-Bonf in case of normal distribution while it performed slightly better in other distributions (t and lognormal). For the test power, the difference in power between $E_g$-BH and W was marginally small while the power of $E_g$-BH, $E_g$-Bonf and W were performed much better than ANOVA in non-symmetric heavy tailed distributions especially lognormal distribution.

The adjusted p values chart has many advantages (a) stretches visible shape to decision maker to gives complete picture and deep insight where the differences happen, keep the level of significance and power of the test similar to original Waerden test and does not need to pairwise comparisons but it gives a glimpse of various information that completes the pairwise comparisons.